\documentclass[aps,prl,twocolumn,superscriptaddress,showpacs]{revtex4}
\usepackage{graphicx}
 \usepackage{psfrag}
\usepackage{ae}
\usepackage{amsmath,amssymb}

\begin{document}
\date{\today}

\title{Polarizable ions at interfaces}

\author{Yan Levin}
\affiliation{Instituto de F\'{\i}sica, Universidade Federal do Rio Grande do Sul, Caixa Postal 15051, CEP 91501-970, 
Porto Alegre, RS, Brazil}

\begin{abstract}
A non-perturbative theory is presented which allows to calculate the solvation free energy of 
polarizable ions near a water-vapor and water-oil 
interfaces.  The theory predicts that larger halogen anions are adsorbed at the interface, while the alkali metal 
cations are repelled from it.  The density profiles calculated theoretically 
are similar to the ones obtained using the molecular dynamics
simulations with polarizable force fields.  

\end{abstract}


\maketitle

There are a number of long standing mysteries in the fields of physical chemistry and biophysics. The Hofmeister
effect~\cite{Ho88}, which has now been known for over $120$ years is, perhaps, one of the oldest and most puzzling ones.   
Hofmeister observed that
different ions have very different effect on stability of protein solutions.  While some electrolytes are very  
efficient at salting-out proteins, others lead to protein precipitation only at much larger concentrations.  A related
mystery, which is also very old, has to do with the surface tensions.  Some hundred years ago
Heydweiller~\cite{He10} noted that adding a strong electrolyte to water leads to increase in the  
surface tension of the water-air interface.
While the dependence on the type of cation is weak, there is a strong variation of the excess surface
tension with the type of anion --- the lighter halides lead to larger  excess surface tension than the heavier ones.  
The sequence is precisely the reverse of the Hofmeister one.  Both effects are completely unaccounted for by the current theories
of electrolytes, which go back to the pioneering work of Debye and Hückel (DH)~\cite{DeHu23,Le02}.  

Application of the DH theory to the study of interfacial properties of electrolyte solutions 
was initiated by Wagner~\cite{Wa24} and
continued by Onsager and Samaras (OS)~\cite{OnSa34}.  These approaches were based on the observation 
that the image charge induced at the air-water
interface repels ions from the surface,
creating a depletion layer.  The Gibbs adsorption isotherm then leads to the
conclusion that
the surface tension of aqueous electrolytes must be higher than that of pure water. Unfortunately, there is no
way to account for ionic specificity within these theories.  Since the hydrated size of all halide ions is nearly
the same --- and this is the only parameter that enters into DH theory --- the OS approach 
predicts that the surface excess should be independent of the type of ion.  
Recently there have been proposed some other approaches, but none have 
proven completely satisfactory~\cite{Le00}.  

Some clues to the failure of the DH and the OS theories started to appeared in the 1990s when the photoelectron emission 
experiments~\cite{MaGi91}
and molecular dynamics simulations with polarizable force fields showed that contrary to the common wisdom, there were ions 
present at the air-water interface~\cite{DaSm93,StBe99}. 
The simulations found that while hard alkali metal ions such as Potassium and Sodium and small halides such as Fluoride~\cite{Br08}
are repelled from the interface, the softer more polarizable anions such as Bromide and Iodine are 
actually attracted to it~\cite{JuTo02}.  
Presence of highly reactive halogens
at the air-water interface of aerosol particles  
might  help to explain the excessive rate of ozone depletion~\cite{KnLa00}. 

In this paper a new class of electrolyte models will be introduced. Unlike the previous approaches,
the polarizability of ions will be explicitly taken into account. 
The new theory is intrinsically
non-perturbative  --- all the moments of the ionic charge distribution, 
and not just the dipole, are taken into account.   The calculated solvation free energies are used to obtain the
interaction potential of polarizable ions with an interface and to   
calculate the anion and cation density profiles.

Since the pioneering work of Debye and Hückel, ions have been modeled as hard spheres with a point charge located
at the center~\cite{Le02}.  While perfectly reasonable for describing bulk properties of electrolytes, this approach is bound to
fail when applied to polarizable ions near a dielectric interface.  The reason for this is easily understood by considering
the simplest model of a perfectly polarizable ion idealized as a conducting spherical shell 
with a mobile surface charge.  When such an ion  moves across a dielectric air (oil)-water interface, the charge on its 
surface shifts progressively from the exposed air/oil portion to the part that still remains hydrated.  
For perfectly polarizable ions, energy cost of charge localization is very low
and is easily compensated by the decrease in the cavitational energy~\cite{RaTr05} as 
the ion moves across the interface.

To make the discussion quantitative,
consider a polarizable ion ---  modeled as a conducting sphere of radius $a$ and charge $q$, see Fig. 1 --- 
at an air (oil)-water interface. Both half-spaces will be treated as dielectric continuums with permittivities
$\epsilon_w$ and  $\epsilon_o$ for water and air (oil), respectively. 
\begin{figure}
\begin{center}
\psfrag{a}{$a$}
\psfrag{th}{$\theta$}
\psfrag{h}{$h$}
\psfrag{z}{$z$}
\psfrag{eo}{$\epsilon_o$}
\psfrag{ew}{$\epsilon_w$}
\psfrag{rh}{$\rho$}
\psfrag{rhm}{$\rho_m$}
\includegraphics[width=4cm]{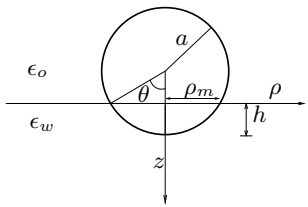}
\end{center}
\caption{Ion of radius $a$ at an interface}
\label{fig1}
\end{figure}
To gain insight into the problem we first consider an ion with one of its hemispheres submerged in water
and the other exposed to air, $h=a$, Fig. 1.  This  problem can be solved exactly, yielding a purely radial electric field
and the electrostatic self energy  of the ion given by
\begin{equation}
\label{e1}
U_s(a)=\frac{q^2}{(\epsilon_w+\epsilon_o) a}\;.
\end{equation}
We can also calculate the ratio of the surface charge 
on the two hemispheres, obtaining $q_o/q_w=\epsilon_o/\epsilon_w$.  This means that for an air-water
interface with $\epsilon_o/\epsilon_w=1/80$, the fraction of the ionic charge located in air is only  $1\%$ of the
total charge!  For a perfectly
polarizable ion half exposed to air, $99\%$ of its charge remains hydrated!  This is very different
from non-polarizable ions, which under the same conditions 
will have  half of their charge exposed to the low dielectric environment, 
at a huge electrostatic self energy cost.  This is the
reason why non-polarizable ions can not penetrate into the interfacial region.  Evidently this is not the case 
for polarizable ions
which can adjust their charge distribution to minimize the electrostatic self energy cost.  
Unfortunately, once we leave the symmetric case of an ion located half way across the interface, 
no exact solution is  possible and approximate methods must be used.  

The self energy of a perfectly polarizable 
ion with its lower extreme situated  a distance $h$ from the interface, see Fig. 1, can be written as,   
\begin{equation}
\label{e2}
U_s(h)=\frac{q^2}{2 \epsilon_w C}\;,
\end{equation}
where the capacitance  $C=a f(\epsilon_o/\epsilon_w, h/a)$ and $f(x,y)$ is a scaling function.  In the limit $\epsilon_o \ll \epsilon_w$,
we can expand $f$ in powers of  $\epsilon_o/\epsilon_w$,
\begin{equation}
\label{e3}
f(\frac{\epsilon_o}{\epsilon_w}, \frac{h}{a})=f(0,\frac{h}{a})+\frac{\epsilon_o}{\epsilon_w} f_1(\frac{h}{a})\;.
\end{equation}
The value of the scaling function $f(0,h/a)$ determines the capacitance in the limit of vanishing air permittivity.  
In this limit the electric field lines originating on the charge inside water must be tangential to the interface, 
so that the normal
component of the electric field vanishes.  Even this
problem, however, is difficult to solve analytically.  Exact  solution is possible, however, 
when $h/a \ll 1$.  In this limit all charge is confined to a small spherical
cap located inside  water.  The curvature effects can be neglected, and  the problem reduces to finding the solution of
a mixed boundary value problem in cylindrical coordinates: $\nabla^2 \phi(z,\rho)=0$, $\phi'(0, \rho)=0$ for
$\rho>\rho_m$, and $\phi(0,\rho)=V$ for $\rho \le \rho_m$, where prime refers to the derivative with respect to $z$,
$V$ is the electrostatic potential of the spherical cap, and $\rho_m$ is its radius, see Fig. 1. 
Mixed boundary value problems are notoriously difficult to study.  Fortunately, this particular one 
can be solved analytically using 
the Hankel transform techniques~\cite{Ja99}.   We find
\begin{equation}
\label{e4}
\phi(z,\rho)=\frac{2 V}{\pi} \int_0^\infty {\rm d}k \frac{\sin(k \rho_m)}{k} J_0(k\,\rho) e^{-k\,z} \;. \\
\end{equation}
The capacitance of the spherical cap can now be calculated to be $C_c=\rho_m/\pi$.  
We note that this is just half the value of the capacitance
of a disk of radius $\rho_m$ in vacuum.  This result can be understood by considering a charged disk 
in front of a dielectric
medium of very low permittivity.  The image charge induced on the interface 
will then be of the same sign as on the disk, and in the limit  
$\epsilon_o \ll \epsilon_w$ it will also be the same in magnitude.  Thus, one needs only half the
charge of the disk in vacuum to have the same potential.  

In view of the natural symmetry of the problem it is  convenient to express  $C_c$ in terms
of the angle variable $\theta$, so that $C_c=a \theta/\pi$ and  $f(0,h/a)=\theta/\pi$.  Writing the capacitance in terms
of $\theta$ 
extend the range of validity of $C_c$ outside the limit of completely flat disc to larger spherical caps. 
In particular, for a
particle  located
half way across the interface, $h=a$, $\theta=\pi/2$, we find $f(0,1)=1/2$, 
which agrees precisely with the exact result of  Eq.~(\ref{e1}). 
Furthermore, comparing Eqs.~(\ref{e2}) and (\ref{e3}) with Eq.~(\ref{e1}), we see that
$f_1(1)=1/2$.  We are now in  position to write an approximate expression for the self energy of a perfectly 
polarizable ion moving 
across a dielectric interface,
\begin{equation}
\label{e5}
U_s(h)=\frac{q^2}{2 \epsilon_w a}\frac{1} {\frac{\theta(h)}{\pi} +\frac{\epsilon_o}{2 \epsilon_w}}\;.
\end{equation}
where $\theta(h)=Re[\arccos(1-\frac{h}{a})]$.  The real part of $\arccos(x)$ is used in order to continue its
validity into the regions $h>2a$, where $\theta(h)=\pi$.  For $h>2a$  the electrostatic self energy 
reduces to $U_s \approx q^2/2\epsilon_w a$, which is the usual Born self energy of a bulk ion.  
In writing Eq.~(\ref{e5}), we have approximated the scaling function $f_1(x)$
by a constant, $f_1(x)=1/2$. This is permissible, since when the ratio $\epsilon_o/\epsilon_w \ll 1$, 
the prefactor of $f_1$ is very small
and the precise value of  $f_1(x)$ is 
not important --- it is completely dominated by the first term of Eq.~(\ref{e3}).
We should note, however, that
although Eq.~(\ref{e5}) is very accurate for $\pi \epsilon_o/(2 \epsilon_w) < \theta < 3 \pi/4$, and for $h/a \gg 1$, 
it does not describe perfectly the crossover from the interfacial to the bulk regime.  The reason for this  
is that  Eq.~(\ref{e5}) does not fully account for the image contribution to the electrostatic energy  
at distances $h>a$. It is possible to
include this corrections into the theory at the expense of more complicated expressions.  
In practice,  however, we note
that the image  contribution is screened very strongly~\cite{OnSa34}, with the characteristic 
length equal to half the Debye length $\xi_D$, 
$U_{im}(z) \approx q^2\exp(-2 z/\xi_D)/(4 \epsilon_w z)$.  Therefore, for concentrations of electrolyte above physiological 
ones $150$mM, the image contribution
decays to zero after only a few Angstroms.  For now, we shall, therefore, ignore the image effect in the crossover region.

It is important to stress the fundamental difference between Eq.(\ref{e5}) and a similar expression for non-polar hard ions. 
If a hard ion is located half way across the interface, it will have half of its charge
exposed to the low dielectric environment.  The electrostatic self energy cost for this is 
approximately $\sim q^2/(\epsilon_o a)$. This is almost two orders of magnitude
larger than the energy cost for a polarizable ion to be at the same position!  This is the reason why non-polarizable
ions will not be found at the interface. 

The force that drives ions towards the interface arises from the cavitational energy.  Presence of ions 
disturbs the hydrogen bond network of water and costs energy.  We can estimate this energy cost  by considering a cavity which
must be  formed
in water to accommodate an ion.  
For small cavities of radius $a<4$ \AA$,$  which do not significantly perturb the hydrogen bonds, 
the energy cost scales with the volume of the void, while for larger
cavities the energy cost scales with the cavity surface area~\cite{LuCh99}.  
This is, the so called, hydrophobic crossover from small to large length scales~\cite{Ch02}.  
Small alkali metal and  halogen ions are  
in the volumetric scaling regime.  If one part of an ion leaves the aqueous environment, the cavitational 
energy will decrease proportionately to the volume exposed.  This results
in a short range cavitation potential which forces ions to move into the air (oil) phase,
\begin{eqnarray}
U_{cav}(h)=\left\{
\begin{array}{l}
 \nu a^3 \text{        for        } h \ge 2 a \nonumber \\
 \frac{1}{4} \nu a^3  \left(\frac{h}{a}\right)^2 \left(3-\frac{h}{a}\right)          \text{        for        } 0<h<2 a
\end{array}
\right.
\end{eqnarray}
From the results of bulk simulations~\cite{RaTr05}, we calculate that $\nu \approx 0.3 k_B T/$\AA$^3$. 
To account for the fact that a cavity containing an ion should be somewhat
larger than the ionic crystallographic size, 
we will use  $\nu \approx 0.5 k_B T/$\AA$^3$. This corresponds to the cavity 
radius about $20\%$ larger than the crystallographic radius.
 
For small non-polarizable ions, the cavitational energy is not sufficient to force 
ions into the low dielectric environment ---
the electrostatic energy cost is way too large.   
On the other hand,  for soft polarizable ions,  the electrostatic self energy cost is very small, 
since the ionic charge distribution can easily deform to remain mostly within the hydrated portion of the
ion.   The significant gain in the cavitational energy, and the low cost in electrostatic self energy,
makes it energetically favorable for polarizable ions to move into the interfacial region.  
The amphiphilic nature of large ions,  such as  hexafluorophosphate $PF_6^-$,
has been known for a long time. The cavitational energy for these ions is so large, 
that they actually adsorb to the interface,  
lowering its surface tension~\cite{Ra57}.  What has been
discovered recently is that smaller polarizable ions apparently can also have some amphiphilic activity --- 
although not sufficiently large to lower the interfacial tension~\cite{JuTo02,MaGi91}. 

So far we have considered only perfectly polarizable ions.  Real ions, however, have finite polarizability.  It is 
not obvious how the effects of finite polarizability can be included in the the formalism developed above. 
In fact it is not even clear,
if the concept of bulk polarizability, as a liner response to the external field, is relevant in the interfacial geometry, 
where a rapid variation of the dielectric constant makes 
all the moments of the charge distribution ---  not just the dipole --- relevant.  
For perfectly polarizable ions we have avoided
this difficulty by solving the full electrostatics problem for a conducting sphere. For ions of finite polarizability,
to have a completely quantitative picture it might be necessary  to go to full ab initio calculations.  
In the absence of such, we can still gain some insight into this difficult problem by considering
a simple model.  In the spirit of Landau, we will construct the polarization energy $U_p$ by exploiting the
symmetries of the problem.  

Consider an ion of radius $a$ and bulk polarizability $\gamma$.  We will define the  relative polarizability
of this ion as $\alpha \equiv \gamma/\gamma_0$, where $\gamma_0=a^3$ is the polarizability of an ideal ion of the same radius,  
modeled as a conducting
sphere. For non-ideal  
ions with $0\le\alpha<1$, the surface charge can not fully adjust to the external electric field.
Therefore, for such ions, there is an
additional non-electrostatic  --- quantum mechanical --- energy cost for 
dislocating ionic charge from its position of equilibrium.  
Suppose that for a given ion
the fraction of its total charge inside water is $x$, then the charge exposed to air (oil) will be $(1-x)q$.  
For ions which are
highly polarizable $x\approx 1$, as long as $\theta$ is not too small . There is, however,
a polarization energy cost for shifting a fraction of the ionic 
charge (assumed to be originally uniformly distributed along the surface of the ion) from its equilibrium position in 
the air portion of the ion to 
the water part.  Within our simple dielectric model $U_p$ must 
be invariant under the transformation $q \rightarrow -q$.  It must also be
invariant under the transformation $\theta \rightarrow \pi-\theta$, 
when $\epsilon_o \leftrightarrow \epsilon_w$, and $x \rightarrow 1-x$. 
To respect these symmetries, the polarization energy must be an even function of the difference 
between the initial (before exposure) and the final (after exposure) 
amount of charge on the part of the ion exposed to air/oil. Taking all these considerations into account and 
recalling Eq.~(\ref{e5}) for a perfectly conducting sphere,
we arrive at the polarization energy for a non-ideal ion, 
\begin{eqnarray}
\label{e7}
\beta U_p(h;x)&=&\frac{\lambda_B}{2 a} \left[\frac{\pi x^2 }{\theta}+ \frac{ \pi (1-x)^2 \epsilon_w}{\epsilon_o (\pi-\theta)} \right] \nonumber \\
&+& g \left[x-\frac{1-\cos(\theta)}{2} \right]^2
\;,
\end{eqnarray}
where $\beta=1/k_B T$ and $\lambda_B=q^2/\epsilon_w  k_B T$ is the Bjerrum length in water.
The terms in the first square brackets of Eq.~(\ref{e7}) are the
electrostatic self energy costs
of the parts of the ion exposed to the water and the air, respectively.  The second brackets contain the
energy cost arising from the induced inhomogeneity in the ionic surface charge distribution. 
The coupling constant $g$ must be a function of
the relative polarizability, $g(\alpha)$. In the limit $\alpha \rightarrow 1$, ion becomes perfectly polarizable,
so that  $g \rightarrow 0$; while in the limit $\alpha \rightarrow 0$, ion becomes completely hard and 
$g \rightarrow \infty$.
To account for these, we will write $g(\alpha)=\chi (1-\alpha)/\alpha$, where $\chi$ is a pure number.  The precise
value of $\chi$ can only be obtained from the ab initio calculations. 
For now we will take it to be of order unity, $\chi \approx 1$.
Minimizing  Eq.~(\ref{e7}) with respect to $x$ gives the fraction of the total charge 
located on the hydrated part of the ion, $x_{min}(h)$.
Substituting this back into Eq.~(\ref{e7}), yields the polarization potential 
that an ion feels as it moves across the interface $U_p(h)=U_p(h;x_{min}(h))$.  
For an ideal ion of  $\alpha=1$ located
half way across the interface $h=a$, the energy $U_p(a)$ reduces precisely to the expression given by Eq.~(\ref{e1}). 
Therefore,  for this case, 
the  formalism developed above gives the exact result. The total solvation potential felt by an ion of 
relative polarizability $\alpha$ is $U_{sol}(h;\alpha)=U_{cav}(h)+U_p(h)$.  Using this potential, we can calculate
the ionic density profiles inside a small water droplet of radius $R$ by explicitly solving a modified 
Poisson-Boltzmann equation,
\begin{eqnarray}
\label{e8}
\nabla^2 \phi(r)&=&\frac{q N}{\epsilon_w} \left(\frac{e^{\beta q \phi(r) -\beta U_{sol}(h;\alpha_a)}}
{\int_0^R r^2\, dr \, e^{\beta q \phi(r) -\beta U_{sol}(h;\alpha_a)}}-\right. \nonumber \\
&&\left. \frac{ e^{-\beta q \phi(r) -
\beta U_{sol}(h;\alpha_c)}}{\int_0^R r^2\, dr\, e^{-\beta q \phi(r) -\beta U_{sol}(h;\alpha_a)}}\right),
\end{eqnarray}
where $r$ is the distance measured from the center of the water droplet, $h=R-r$, 
and $\alpha_c$ and $\alpha_a$ are the relative
polarizabilities of cations and anions, respectively.
\begin{figure*}[htb]
\begin{center}
\psfrag{rh}{\Huge $\rho(r)$}
\psfrag{r}{\Huge $r$}
\includegraphics[angle=270,width=5cm]{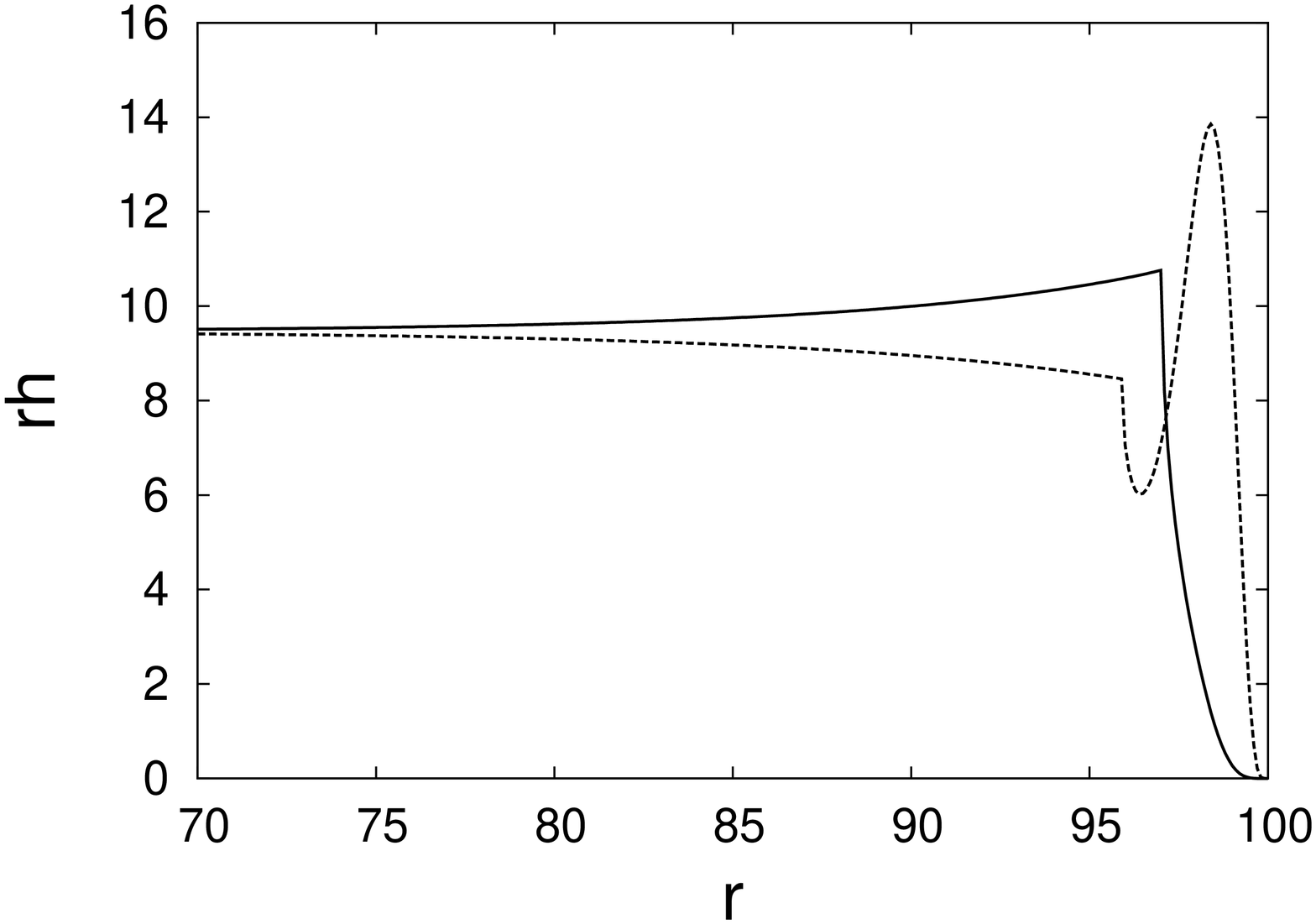}
\includegraphics[angle=270,width=5cm]{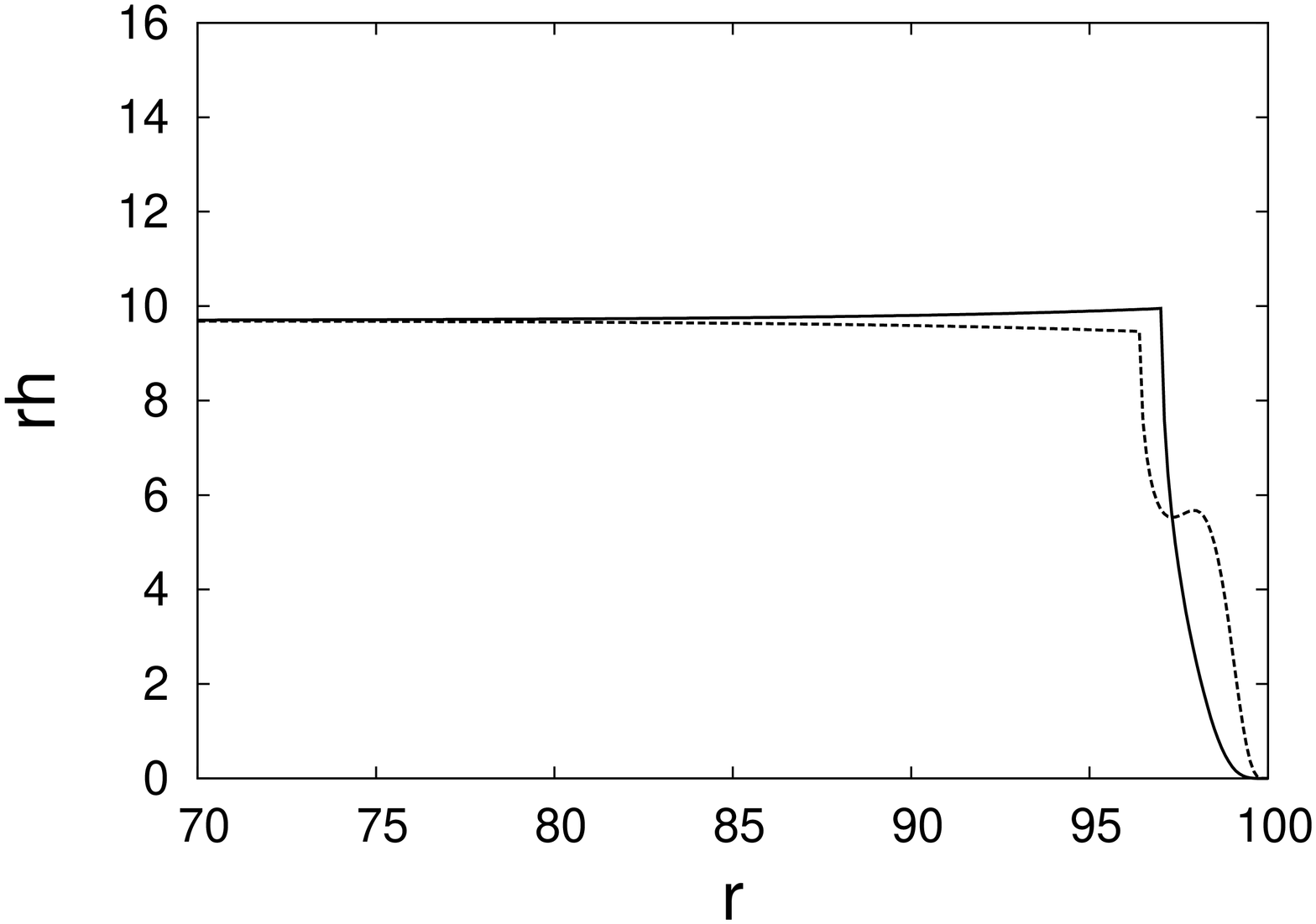}
\includegraphics[angle=270,width=5cm]{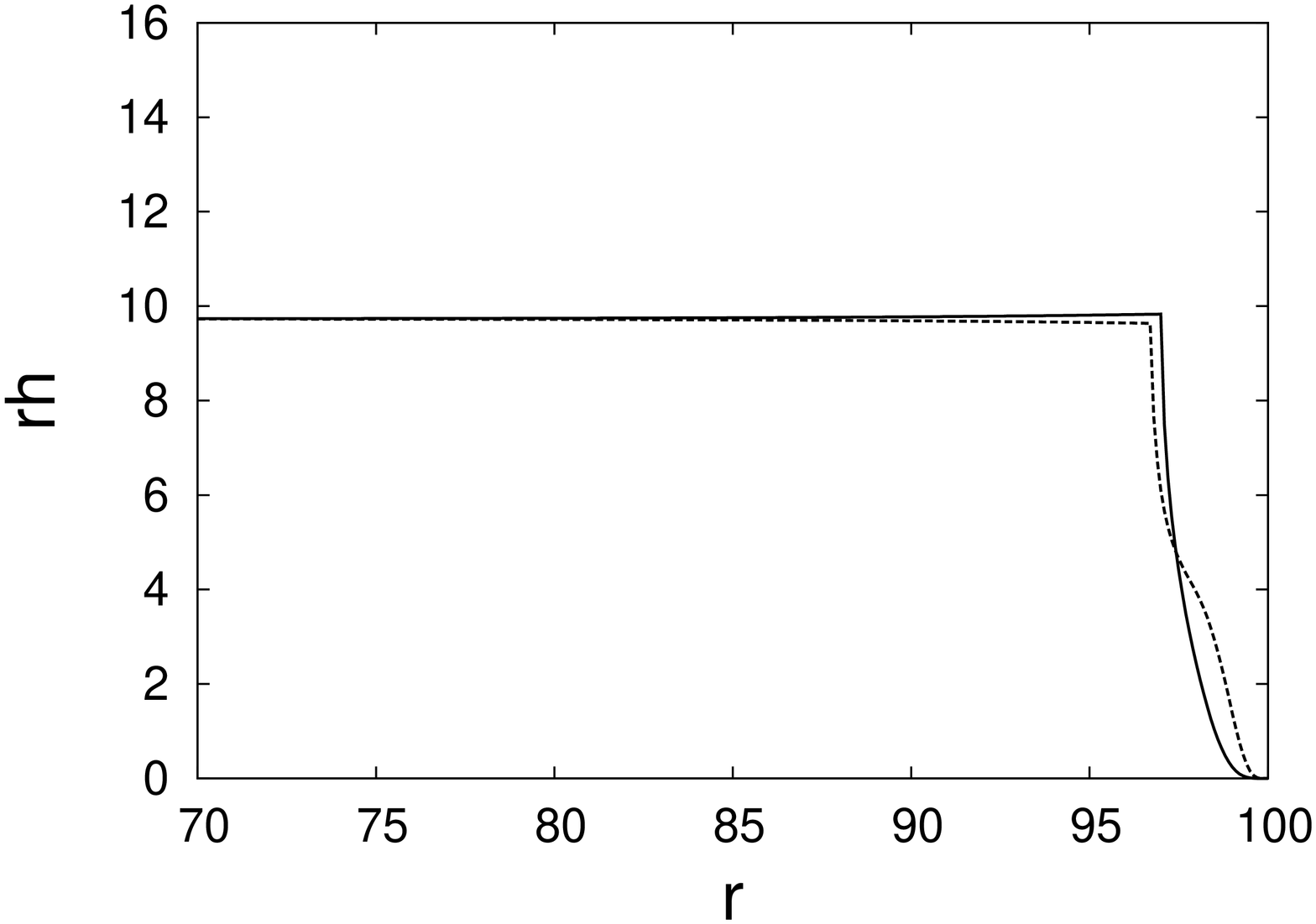}
\end{center}
\caption{Density profiles ($\times 10^5$) of $KI$,$KBr$ and $KCl$, from left to right.  
The radius of the water drop is $R=100$\AA$\,$ and
it contains $N_+=N_-=380$ ions at physiological concentration of $150$mM. Solid curve is for cation ($K$) and the dashed ones are for 
anions. In all cases Potassium 
is depleted from the interface, but there is a formation of a double layer, where the excess of anion directly
at the interface results in a build up of $K$ in its vicinity.}
\label{fig2}
\end{figure*}
The density profiles for electrolyte solutions of $KI$, $KBr$, and $KCl$ are presented in Fig. 2. The polarizabilities of ions
$\gamma_K=0.79$\AA$^3$, $\gamma_I=7.4$\AA$^3$, $\gamma_{Br}=5.07$\AA$^3$, $\gamma_{Cl}=3.77$\AA$^3$, where take from 
Ref. \cite{PyPi92} and
the ionic sizes $a_K=1.49$\AA,  $a_I=2.05$\AA,  $a_{Br}=1.8$\AA,  $a_{Cl}=1.64$\AA,  from Ref. \cite{GuAd60}. 
In agreement with the polarizable 
force fields simulations, the theory predicts that Iodine is strongly adsorbed at the air water-interface. 
We also find that there
is a significant concentration of Bromide, while Chloride,
Potassium and Fluoride (not shown) are depleted from the interfacial region. 
The current theory, however, predicts that there is 
less halide ion adsorption than is found in the simulations.  
The difference might due to the overestimate of the
neat water-vapor surface potential predicted by the polarizable force field simulations to be
$-500$mV$\approx -20k_B T/q$.  Such a huge junction potential will irreversibly drive 
polarizable halides toward the vapor phase, resulting in a large density build up along the interface. 
Recent ab initio simulations~\cite{KaKu08}, however, find a much smaller  
contact potential, $-18$mV$\approx -0.7k_B T/q$, for a water-vapor interface.  This might lead to a 
smaller adsorption, in line with the predictions of the present theory.

There is a subtle interplay between  ionic size and 
polarizability.  Although the absolute value of polarizability for halide ions varies dramatically, the relative 
polarizability $\alpha$ is practically the same for all ions  $\sim 0.85$.  On the other hand, the 
relative polarizability of Potassium is
almost four times lower.
Future work should try to optimize the parameters of the theory $\nu$ and $\chi$ by fitting $U_{sol}$
to the full ab initio simulations.  Unfortunately, no such calculations are available at this  moment. 
Finally, since the surface tension of electrolyte solution is directly related to ionic adsorption in the interfacial region,
the theory developed 
also accounts for the Hofmeister series for halogens.

I am grateful to Felipe Rizzato and Alexandre P. dos Santos for help with numerics. This work is partially supported 
by CNPq and by the US-AFOSR under the grant FA9550-06-1-0345.


\end{document}